\documentclass[a4paper,12pt]{article}

\def\ad   {a^{\dagger}}
\def\cd   {c^{\dagger}}

\def\cd   {c^{\dagger}}

\def\al   {\alpha}
\def\be   {\beta}
\def\bet  {\beta}

\def\del  {\delta}

\def\HH {\hat H}

\def\HQ {\hat Q}

\def\HH {\hat H}

\def\HQ {\hat Q}

\def\CE {{\cal E}}
\def\CL {{\cal L}}

\def\CO {{\cal O}}

\def\CH {{\cal H}}

{\it}
{\it}
\def\expo {\rm {exp}}{\it}
\def\tr {\rm {tr}}{\it}
{\it}
\def\exp {\rm {e}}{\it}
\def\be {\beta}
\def\bet {\beta}

\def\de {\delta}

\def\om {\omega}
\def\Om {\Omega}

\usepackage{times}
\usepackage{graphicx}
\begin{document}
\title{ A Time-Dependent Multi-Determinant approach to nuclear 
dynamics.}
\author{G. Puddu\\
       Dipartimento di Fisica dell'Universita' di Milano,\\
       Via Celoria 16, I-20133 Milano, Italy}
\maketitle
\begin {abstract}
         We propose a Time-Dependent Multi-Determinant approach to the description of the time evolution of the nuclear wave
         functions (TDMD). We use the Dirac variational principle to derive the equations of motion using as ansatz for 
         the  nuclear wave function a linear combination of Slater determinants.
         We prove explicitly that the norm and the energy of the wave function are conserved
         during the time evolution. This approach is a  generalization of the time-dependent
         Hartree-Fock method to many Slater determinants.
         We apply this approach to a case study of ${}^6Li$  using the N3LO interaction renormalized
         to $4$ major harmonic oscillator shells. We solve the TDMD equations of motion using
         Krylov subspace methods of Lanczos type.
         As an application, we discuss  the isoscalar monopole strength function.
\par\noindent
{\bf{Pacs numbers}}: 21.60.-n,$\,\,$  24.10.Cn,$\,\,$ 31.70.Hq
\par\noindent
\vfill
\eject
\end{abstract}
\section{ Introduction.}
     The time-dependent Hartree-Fock method (TDHF) and its quasi-particle generalization, the time
     dependent Hartree-Fock-Bogoliubov method (TDHFB), are central tools in studying nuclear
     dynamics (see for example ref. [1], ref. [2] for a recent review and references in there).
     In these approaches the time dependence of the nuclear wave function is studied 
     under the assumption that the nuclear wave function can be described 
     by a single Slater determinant or by a quasi-particle determinant wave function. Usually
     nuclear excitations, for example giant resonances, are studied in the approximation of small
     amplitude motion around the static solution (RPA or QRPA). In this case, the description of nuclear excitations 
     reduces to the solution of a large eigenvalue problem.
     Despite the enormous matrix dimensions, the RPA or QRPA equations are solved nowadays using efficient
     Krylov projection techniques of Arnoldi type (see for example ref.[3] for recent applications).
     Recently, the time-dependent coupled-cluster method (refs. [4],[5]) has been revisited (ref. [6])
     and it has been applied to light
     nuclei (ref.[7]) using the N3LO interaction (ref. [8]) transformed by the similarity renormalization group.
\par
     In this work we discuss a Time-Dependent Multi-Determinant (TDMD) approach whereby 
     the nuclear wave function
     is approximated by a linear combination of several Slater determinants. This approach is the time dependent version
     of the Hybrid Multi-Determinant (HMD) approach (refs. [9]-[11]).
     Each Slater determinant is built from different single-particle wave functions of the most generic type. 
     To the author knowledge, this approach has never been considered in nuclear physics.
     In this sense, this is an exploratory study. Our starting point is the Dirac variational principle
     which, as well known, leads to the time-dependent Schroedinger equation in the most general case,
     or to the  TDHF equations (ref.[12]) if the  nuclear wave function is approximated  by a single Slater determinant.
     Using the Dirac variational principle, we derive the equations of motion and prove explicitly
     that the time evolution conserves the norm and the energy of the wave function. The equations of motion
     for the single-particle wave functions are of the type $ i L \dot\psi = R$ where $R$ is an 
      energy gradient,
     $\psi$ is the set of single-particle wave functions of all Slater determinants,  
     and $L$ is a matrix of  large dimension related to
     the time derivative of the norm of the wave function (which will be discussed in detail below). The
     actual evaluation of the wave function as a function of time is performed using the
     Direct Lanczos method (DL) for the solution of a  large linear system. The DL method belongs to the
     family of Krylov subspace  methods for the solution of linear systems (an excellent review of
      these methods can be found
     for example in ref. [13]). These methods for eigenvalue problems include the familiar 
     Lanczos method
     used in the shell model approach to nuclear structure (refs. [14],[15]) and the Arnoldi method used in solving the RPA or
     QRPA eigenvalue problems (ref.[3]). The basic idea of these methods is the following. Although we may not be able 
     to store a matrix (e.g. the nuclear Hamiltonian matrix) we can easily evaluate the matrix to vector product.
     In our case, although $L$ is not as large as the shell model Hamiltonian matrix, it can hardly be stored except
     in simple cases. However the matrix to vector product appearing in the equations of motion
     is trivial to evaluate, and  the Laczos method is the ideal one. We solve the equations of motion, as an  
     exploratory study, in the case of ${}^6Li$ using the N3LO interaction renormalized to $4$ major
     oscillator shells with the Lee-Suzuki (ref.[16],[17]) method, in order to reduce the otherwise 
     very large single-particle space. We use the time-dependent wave function  obtained in this way to evaluate
     strength functions.
     Our ultimate goal is to extend ab-initio methods to time-dependent problems, such as the evaluation
     of strength functions, starting from a two-body nucleon-nucleon interaction.
\par 
     Our approach is different from the Multiconfiguration Time-Dependent Hartree (or Hartree-Fock) method
     (MCTDHF)  used in quantum chemistry (ref. [18]-[20]). The MCTDHF is a time dependent version
     of the shell model written in the full Hilbert space. The MCTDHF method uses a time-dependent
     linear combination of all possible Slater determinants. The time-dependent coefficient of such a 
     linear combination is a function of all possible many-body configurations and it is obtained using
     the equations of motion. Since the ansatz for the many-body wave function is not unique, one restricts
     the freedom in the many-body wave function by imposing orthogonality among the single-particle wave functions.
     As shown in ref. [19] this amounts to a redefinition of the coefficient of the linear combination.
     The only difference between an exact treatment of the time evolution of the many-body wave function and
     the MCTDHF approach is that in the latter the single-particle basis is time dependent.
     In the MCTDHF approach, at a given value of time, all Slater determinants are built from the same 
     time-dependent single-particle
     basis, that is, each of them is a n-particle-n-hole excitation from the lowest one. 
     In our approach, instead, each Slater determinant is built from a different time-dependent single-particle basis.
     Moreover, we consider several and not all possible Slater determinants and we do not have the freedom
     of imposing orthogonality between the single-particle wave functions belonging to different Slater determinants. 
     Rather, we  consider the most 
     generic Slater determinants, in the same spirit of the HMD method. Our approach is not limited
     by the dimension of the Hilbert space. Each Slater determinant, in our approach, is equivalent to a
     rather large number of linear combinations of the Slater determinants of the MCTDHF approach. As a consequence,
     the equations of motion in the MCTDHF approach are different from the ones of the TDMD  approach  (cf. ref. [18]-[20]
      and section 2a of this work).
\par
     The outline of this paper is as follows. In section 2 we derive the equations of motion in the TDMD approach
     using the Dirac variational principle, we prove that these equations of motion
     conserve the norm and the energy of the nuclear wave function and discuss how to fix uniquely the
     solution of the equations of motion for the single-particle wave functions. We also briefly discuss the
     imaginary time version of these equations of motion. At the end of section 2 we discuss the
     'static' solutions of these equations and show that the time propagation of these solutions generates
     a time-dependent phase factor common to all Slater determinants (in some sense this  is the generalization of
     the single-particle energies), In section 3 we discuss the   numerical
     method and in section 4 we discuss the application of our method to the nuclear strength function using the boost
     method in order to determine the excitation spectrum. 
\bigskip
\bigskip
\section{ The time-dependent variational principle.}
\bigskip
\par
{\it {2a. Equations of motion and conservation laws.}}
\bigskip
\par
      The Dirac time-dependent variational principle states that the time evolution of the nuclear
      wave function is obtained by varying the action
$$
S_1 =\int_{t_1}^{t_2} dt \CL_1=\int_{t_1}^{t_2} dt [ i\hbar < \psi|\dot\psi>- <\psi|\HH|\psi>]
\eqno(1a)
$$
     or equivalently
$$
S_2 =\int_{t_1}^{t_2} dt \CL_2=\int_{t_1}^{t_2} dt [-i\hbar < \dot \psi|\psi>- <\psi|\HH|\psi>]
\eqno(1b)
$$
     with respect to $|\psi>$ and $< \psi|$ independently, under the constraint that the wave function
     is held fixed at the initial and final times $t_1$ and $t_2$.  The use of the most general wave function
     in the Hilbert space reproduces  the Schrodinger equation and its complex conjugate.
     The TDHF approximation is obtained if the wave function is approximated by  one Slater determinant.
     In what follows we drop $\hbar$ with the understanding that the unit of time is 
     $1 MeV^{-1}\simeq 6.6 \times 10^{-22}sec$.
     We consider the following ansatz for the nuclear wave function
$$
|\psi>=\sum_{S=1}^{N_w} |U_S>
\eqno(2)
$$
     where $|U_S>$ is a Slater determinant and $N_w$ is their  number.
     These Slater determinants for $A$ particles are of the most generic type and are written as
$$
|U_S>= \cd_{1S}\cd_{2S}...\cd_{AS} |0>
\eqno(3)
$$
     $A$ being the number of particles, $S$ labels the Slater determinant and
$$
\cd_{\al S}= \sum_{i=1,N_s} U_{i,\al S } \ad_i,\;\;\;(\al=1,2,..,A)
\eqno(4)
$$
     in the above equation, $\ad_i$ is the creation operator in the single-particle (e.g. harmonic
     oscillator) state $i$, $N_s$ is the number of the single-particle states
     and $U$ is the single-particle wave function in the h.o. representation.
     Note that these single-particle wave functions are different for each Slater determinant 
     labeled by the index $S$.
     In what follows we label particles with greek letters and single-particle states with latin letters.
\par
     As mentioned in the introduction, in the MCTDHF (cf. ref.[20]) each Slater determinant
     is written  as a  multi-particle multi-hole excitation built on the first one. Moreover in the
     MCTDHF approach it is essential to multiply each Slater determinant by a time-dependent amplitude
     and the time-dependent single-particle states can be taken orthogonal to each other.
     That is, in the MCTDHF approach, $ |\psi>=\sum_{[n_1,n_2..]} A(n_1,n_2,..,t) |n_1,n_2,..t>$,
     with the sum extending over all possible allowed values of the occupation numbers of the
     time-dependent basis, i.e. $n_1,n_2,..=0,1$.
     These considerations illustrate the basic difference between the TDMD approach proposed in this work 
     and the MCTDHF approach.
\par
     We assume that  each Slater determinant
     is a product of a neutron and a proton Slater determinant.
     The ansatz
     (3)-(4) is the same of the Hybrid multi-determinant method (refs.[9]-[11]) used in variational 
     calculations. Usually in the HMD method,
     a projector to good quantum numbers (angular momentum and parity) is applied to the wave function of eq.(2),
     in order to decrease the otherwise  large number of Slater determinants needed to obtain accurate energies, for example
     for the yrast states. In this work, we do not use
     projectors to good quantum numbers. We do this  in order to simplify
     the equations and the proof of the conservation of the energy and of the norm.
     The Slater determinants are not orthogonal to each other and are 'deformed', that is, they
     do not have good quantum numbers. At the initial time they could be the result of a partially
     converged variational calculation as given by the HMD method, or converged variational wave functions
     'boosted' by some excitation operator (e.g. dipole, quadrupole , etc.). 
     We do not have the freedom  to impose the
     orthogonality between the single-particle wave functions belonging to different Slater determinants,
     although we can impose orthogonality between the single-particle wave functions of the same Slater
     determinants.
\par
     Although the Dirac variational principle determines uniquely the time dependence of
     the Slater determinants, it
     does not uniquely fix the single-particle wave functions $U_{i\al S} $. In order to see this,
     let us perform the following transformation of the generalized creation operators
     defined in eqs.(3) and (4). 
$$
\cd_{\al S}= \sum_{\bet=1}^A g_{\al,\bet}(S)\cd_{\bet S}{'} \;\;\;\;(\al=1,2,..,A)
\eqno(5)
$$
     for every $S$.
     In other words, we mix the particle labels in each Slater determinant, but we do not mix
     the particle labels of different Slater determinants. Each Slater determinant can be rewritten as
$$
|U(S)>= \det(g(S)) \cd_{1 S}{'} \cd_{2 S}{'} ...\cd_{A S}{'} |0>
\eqno(6)
$$
     Therefore, provided $\det(g(S))=1$, the same Slater determinant can be obtained using
     the new generalized creation operators
$$
\cd_{\al S}{'}=\sum_i \ad_i U'_{i \al S}
\eqno(7)
$$
      with. in matrix notation,
$$
U'(S)=U(S) \tilde g(S) ^{-1}
\eqno(8)
$$
     Hence, if the $U$'s are a solution of the equations of motion (discussed below)
     also the $U'$ given by equation (8) with any $g$ (provided $\det(g)=1$), 
     will satisfy the same equations of motion. This kind of gauge 
     invariance implies that the equations of motion, although they determine the
     time evolution of the set of Slater determinants, they do not determine unambiguously the
     time evolution of the single-particle wave functions $U(S)$. Since $g$ is arbitrary (provided 
     $\det(g)=1$), we have
     $A^2-1$ free parameters for each Slater determinant. In order to uniquely specify 
     the solutions of the equations of motion we select the matrix $g$ such that
$$
U_{\al \bet}'= diag(1,1,...,U'_{AA})_{\al \bet}\;\;\;\;\;(\al,\bet=1,2,..A)
\eqno(9)
$$
    for the $A\times A$ submatrix of $U$ for each Slater determinant.
    In eq. (9), $U'_{AA}$ is the determinant of the $A\times A$ submatrix of $U$
    This point will be further discussed after the equations
    of motion have been derived.
    We assume that all Slater determinants have been recast so that the $A\times A$ submatrices
    of the single-particle wave functions are as in eq.(9) and in what follows we shall
    drop the prime. In this way we effectively decrease the number of unknowns.
\par
     We now proceed to determine the equations of motion of the single-particle wave functions $U$.
     In what follows, since we always have pairs of indices $S$ and $S'$, the Slater determinant $|U_S>$ will have
     the label $S$ (even though sometimes it will be omitted) and the  the complex conjugates of $|U_{S'}>$,
     $<0|c_{AS'}...c_{1S'}$, where
$$
c_{\al,S'}= \sum_i  V_{\al,i S'} a_i
\eqno(10)
$$
     will have the label $S'$. 
     $V_{S'}$ is the Hermitian conjugate of the matrix $U_{S'}$. We do this in order to use simple matrix notations,
     and to avoid confusion between $U$ and $U^{\dagger}$ for different $S$ and $S'$ since often
     we omit the labels $S$ and $S'$ in order to shorten  the equations.
     The Dirac variational principle gives (the bra will be denoted as $<V|$ ) 
$$
i \sum_{S} \de_{V(S')} < V_{S'}| \dot U_S> = \de_{V(S')} \sum_S <V_{S'} |\HH| U_S>
\eqno(11a)
$$
$$
-i \sum_{S'} \de_{U(S)} <\dot V_{S'}|U_S> = \de_{U(S)} \sum_{S'} <V_{S'} |\HH| U_S>
\eqno(11b)
$$
     where we have shown explicitly the quantities which are varied.
     In what follows, we quote the results for the overlaps and for the matrix elements
     of the Hamiltonian (cf. ref.[9]).
     The Hamiltonian is
$$
\HH= {1\over 2}\sum_{ijkl} H_{ijkl}\ad_i\ad_j a_l a_k
\eqno(12)
$$
     where we  recast  the one-body term into the two-body interaction, as done in shell
     model calculations. The matrix elements of $H$ are antisymmetrized (i.e. $H_{ijkl}=-H_{ijlk}$).
     For any $V$ and $U$, (relative to the Slater determinants $S'$ and $S$ respectively) 
      let us define
$$
G = (VU)^{-1},\;\;\; W = G V,\;\; X=UG,\;\;\;\rho= UGV,\;\;\; F = 1-\rho
\eqno(13)
$$
     The matrix $G$ has indices $\al,\bet=1,2,..,A$. The matrix $W$ has indices $\al,i$,
     the matrix $X$ has indices $i,\al$ while $\rho$ and $F$ have indices $i,j=1,2,..,N_s$.
     The matrix $\rho$ is the generalization of the density matrix in TDHF and satisfies
     the relations $\tr\rho=A$ and $\rho^2=\rho$ for any $S'$ and $S$, as it can easily be verified.
     We have then (cf. ref.[9]) 
$$
< V |U>= \det(VU)
\eqno(14)
$$
$$
<V|\HH|U> = <V|U>  \tr (\Gamma \rho)
\eqno(15)
$$
     where the matrix $\Gamma$ is given by
$$
\Gamma_{ij}=\sum_{pq} H_{piqj}\rho_{qp}
\eqno(16)
$$
     Let us note that the exchange term is the same of the direct since the matrix elements
     are antisymmetrized.
     The equations of motion eqs.(11a),(11b) (EOM1 and EOM2) can be derived using the matrix identity,
     for any matrix $M$,
$$
\de \det (M) = \det (M) \tr(M^{-1} \de M)
\eqno(17)
 $$
     Then it is easy to verify that
$$
<V|\dot U>= <V|U> \tr(G V \dot U)
\eqno(18a)
$$
$$
<\dot V| U>= <V|U> \tr(G\dot V U)
\eqno(18b)
$$
     and that the explicit form for EOM1 is (using the identity $\de M^{-1}=-M^{-1} \de M M^{-1}$) 
$$
i \sum_{r\mu,S} \det(VU)\big ( X_{i \al} W_{\mu r}+F_{ir}G_{\mu\al}\big ) \dot U_{r \mu S} =
\sum_S \det(VU) \big ( X_{i\al}\CE +2 (F\Gamma X)_{i\al})
\eqno(19)
$$
     where $\CE$ is the energy functional
$$
\CE= \tr (\Gamma\rho)
\eqno(20)
$$
     The equation of motion EOM2 can be obtained in the same way and is given by
$$
-i\sum_{\mu,r,S'} \det(VU) \big( W_{\al i} X_{r \mu}+ G_{\al\mu}F_{ri}\big) \dot V_{\mu r S'}=
\sum_{S'} \det(VU) (\CE W_{\al i} +2 (W\Gamma F)_{\al i})
\eqno(21)
$$
      These equations need a few comments. First, if we recast then in a schematic matrix notation
$$
i L^{(1)} \dot U = R^{(1)}
\eqno(22a)
$$
$$
-i L^{(2)} \dot V = R^{(2)}
\eqno(22b)
$$
      the dimension of the linear systems to be solved can be rather large. For example
      In the case of ${}^{24}Mg$ with $7$ major shells $(N_s=168)$ for $10$ Slater determinants,
      the matrix $L$
      is $20160\times20160$ (for both neutrons and protons), for a larger number of major shells 
      or for heavier nuclei, the storage
      of this array in the computer memory can be a problem. Moreover these
       matrices seem to have some kind of separable structure. $L^{(1)}$ for example contains a
      separable term in the indices $(i \al)(\mu r)$  and another separable term  in the
      $(ir)(\mu \al)$ indices. This implies that although we may not be able to store
      the matrix $L$, we can very easily perform the matrix to vector product. We only
      needs to store the matrices $X, W, F$ and $ G$, in the case of $Mg$, of  dimension
      $168\times 12,12\times 168 $ and  $168\times 168$ for every S and S'. 
      These matrices are the same matrices used in the HMD variational calculations.
      In the past few decades, linear systems of this type, for which the matrix cannot
      be stored but  the matrix to vector product can easily be performed, have received
      a major attention in applied mathematics using the so called Krylov subspace
      techniques. These techniques are  precisely of the same kind  one uses in standard 
      shell model calculations (ref.[13]). They will be briefly recalled in the next section.
      A systematic treatment can be found in ref. [13] (note however that in ref. [13] the
      convention for the scalar product  is $<x|y>=\sum x_i y^{\star}_i$).
      Equations of motion EOM1 and EOM2 are equivalent.
      The matrix $L^{(1)}$ and $L^{(2)}$ are Hermitian.  
\par
      One can show  that the norm of the wave function  is preserved during the time evolution, using
      the explicit form of the equations of motion.
      From eqs. (14) and (17) one has
$$
d<\psi|\psi>/dt = \sum_{S S'} <V|U> \tr [G(\dot V U+V \dot U) ]
\eqno(23)
$$
        with the understanding that $S'$ refers to $V$ and $S$ to $U$.
       From EOM1 eq. (19), multiplying by $ V_{\al i S'}$ and summing over the indices one has
$$
i \sum_{S' S} <V|U> [\tr (\rho) \tr (G V\dot U)+ \tr(F\dot U G V)]=
$$
$$
 \sum_{S' S}<V|U> [\tr(\rho) \CE+2 \tr(F\Gamma\rho)]
\eqno(24)
$$
       From EOM2 of eq. (21), multiplying by $U_{i \al S}$ and summing over the indices one has
$$
-i \sum_{S S'} <V|U>[ \tr (\rho) \tr (G \dot V U)+ \tr( X\dot V F)]=
  \sum_{S' S}<V|U>[\tr(\rho) \CE+2 \tr(\rho\Gamma F)]
\eqno(25)
$$
      Subtracting eqs. (24) and (25), and since for any $S S'$, $\;\tr (\rho) =A$, we have
$$
i \sum_{S S'} <V|U> [A\;  \tr[G(\dot V U+V\dot U)]+\tr(F\dot U W+X \dot V F)]=
$$
$$
\sum_{S' S}<V|U> 2 
\tr(F\Gamma\rho-\rho \Gamma F)
\eqno(26)
$$
      The right hand side of this equation is $0$ since $F=1-\rho$.
      Next, since $\tr(\dot \rho)=0$ for any $S'S$, using the definition of $\rho$ given in eq.(13) and
      the cyclic property of the trace,
      we have 
$$
\tr[G(\dot V F U+ V F \dot U)]=0
\eqno(27)
$$
      This equation can also be verified directly using the definitions in eq. (13).
      Hence from eq.(26), using the cyclic property of the trace, one has
$$
i \sum_{S S'} <V|U> \tr[G( \dot V U+V\dot U)]=0
\eqno(28)
$$
      which is the time derivative of the norm (cf. eq.(23)). 
      Next we shall prove that the energy is constant during the time evolution.
      We need to prove that
$$
d <\psi|\HH|\psi>/dt=0
\eqno(29)
$$
      since the norm of the wave function is a constant. Let us set
$$
\CH[V,U]= <\psi|\HH|\psi>,\;\;\;\CO[V,U]= <\psi|\psi>
\eqno(30)
$$
      The Lagrangian associated to EOM1 can be rewritten as
$$
\CL_1=i\sum_a {\partial \CO\over \partial U_a} \dot U_a -\CH
\eqno(31)
$$
      where $a=(i \al S)$ for brevity.
      EOM1 can then be written as
$$
i \sum_a {\partial^2 \CO\over \partial V_b\partial U_a}\dot U_a= {\partial \CH\over \partial V_b}
\eqno(32)
$$
      for all $b=(\bet j S')$. Similarly the Lagrangian associated with EOM2 can be recast as
$$
\CL_2=-i \sum_b {\partial \CO\over \partial V_b}\dot V_b-\CH
\eqno(33)
$$
      and EOM2 can be recast as
$$
-i \sum_b {\partial^2 \CO\over \partial V_b\partial U_a}\dot V_b= {\partial \CH\over \partial U_a}
\eqno(34)
$$
      Multiplying eq.(32) by $ \dot V_b$ and  summing over the indices, and similarly multiplying eq.(34)
      by $\dot U_a$ and summing over indices, after subtracting the two results, we obtain
$$
\sum_a {\partial \CH\over \partial U_a}\dot U_a + \sum_b {\partial \CH\over \partial V_b}\dot V_b =0
\eqno(35)
$$
      which is precisely the time derivative of $\CH$.
      These two constants of motion are a valuable test to check whether the equations of motion
      have been integrated with reasonable accuracy.
\par 
     Before leaving this subsection. let us discuss the consequence of the fact that the physical
     objects are the Slater determinants, rather than the single-particle wave function. Without
     fixing $A^2-1$ coefficients for each Slater determinant, we would have an infinite
     number of solutions for the linear system of eq.(19) or eq.(21) in the unknowns $\dot U$ or 
     $\dot V$. This implies that 
     $\det(L)=0$ and that a direct attempt to solve the equations of motion by matrix
     inversion will fail. We must first fix $A^2-1$ coefficients for each Slater determinant
     before any attempt to use direct methods (such as Gaussian elimination) to solve the linear 
     system. This means that we can consider all $\dot U_{\al,\bet}=0 $ for $\al,\bet=1,..,A$,
     except $\al=\bet=A$, and reduce the dimension of the linear system accordingly.
     The condition of eq.(9) is not equivalent to orthogonality of the single-particle
     wave functions (even for the same Slater determinant). We find eq.(9) simpler to implement
     for several Slater determinants than the orthogonality. as shown by the structure
     of the equations  of motion.
     Only in the case of a single Slater determinant they can be made orthogonal and orthogonality
     is preserved during the time evolution. All these
     considerations have been tested numerically. We did not find any need to enforce eq. (9)
     using Krylov subspace techniques. Actually all initial calculations have been 
     performed without the gauge fixing condition of  eq.(9). Note also that if
     we impose (for a given $S$) orthogonality between the single-particle wave functions
     we would have to introduce Lagrange multipliers, while the condition of eq.(9), simply
     reduces the number of unknowns in the linear system of eq.(19). 
\bigskip
\par
{\it{2b. Imaginary time equations of motion.}}
\par
      Propagation in imaginary time can be used to determine the best approximation to the ground-state
      for a specified number of Slater determinants. As $\tau=i t\rightarrow \infty$ we obtain the 
      ground-state of the system. We solve the following imaginary time equations of motion
$$
L \dot V = - R
\eqno(36)
$$
      where $L$ and $R$ are given in the previous subsection in eq. (21). We consider EOM2 since
      the basic matrices in eq.(13) can be taken from HMD computer programs, which have accurately been tested.
        We also solve the variational
      problem using the HMD method (which is a quasi-newtonian method). The technical details 
      of the variational methods used in the HMD approach can be found in ref. [21].
      The results from the HMD method can be used as initial start
      in eq. (36) and vice versa. We obtain the same energies from the two methods and this is  a  
      strong validation  test of our computer programs. 
      Once $\dot V$ in eq. (36) has be found, we determine $V$ using Runge-Kutta methods with a time interval
      sufficiently small so that the energy decreases as a function of the imaginary time. Typical values
      for the imaginary time interval are $10^{-2}, 10^{-3} MeV^{-1}$. 
\bigskip
\par
{\it{2c. The static solutions.}}
\par
      Let us suppose that we have found the ground state wave function for a selected number of Slater determinants,
      either by imaginary time propagation
      or with the variational HMD method, and let us call these single-particle wave functions  $\overline V(S')$.
      As in the the TDHF approximation, we can propagate in real time these static single-particle wave functions and obtain the
      single-particle energies.  However, in the case of several Slater determinants we cannot define the single-particle
      energies since we do not have a self-consistent eigenvalue problem as in the the HF approximation. The question naturally
      arises whether one can define some type of single-particle energies from the the evolution
      of the static solutions $\overline V(S)$. In the case of several Slater determinants, since we do not impose
      orthogonality between single-particle wave functions, these  can mix. Hence we seek
      solutions of the type
$$
V_{\al i S'}(t) = \sum_{\be=1}^A f_{\al \be}(t,S') \overline V_{\be i S'}
\eqno(37)
$$
      with the $A\times A$ matrix $f$ determined by the equations of motion EOM2. The matrices $U,X,W,G,\rho,\Gamma$ and 
      $F$  for a pair of Slater determinants $S,S'$ obey the relations, in a matrix notation,
$$
U(S) = \overline U(s) f^{\dagger}(t,S),\;\;\; G = f^{\dagger -1}(t,S) \overline G f^{-1}(t,S')
$$
$$
W = f^{\dagger -1}(t,S) \overline W,\;\;\;\;\; X = \overline X f^{-1}(t,S'),\;\;\; \rho=\overline \rho
\eqno(38)
$$
$$
\CE = \overline \CE,\;\;\;\; F=\overline F,\;\;\;\; \Gamma=\overline \Gamma
$$
     Quantities with the overline are  obtained with the imaginary time propagation or with the HMD method.
     The equation EOM2 then  gives
$$
 \sum_{S'} \det(\overline V \overline U) \det(f(t,S')) [ \overline W \tr(\overline X M )+ \overline G M \overline F]=
 \sum_{S'}\det(\overline V \overline U) \det(f(t,S')) [ \overline \CE \overline W +2\overline W\overline \Gamma\overline 
F]
\eqno(39)
$$
     where
$$
M= -i f^{-1}(t,S') \dot f(t,S')
\eqno(40)
$$
     We seek time-independent $M$ i.e. $f=\expo(i M t)$. Since eq.(39) has to be valid at all times 
     $ \det(f(t,S'))$ must be independent of $S'$, i.e. all Slater determinants must evolve with the same
     phase factor $\expo[i \tr( M) t]$. As a consequence the Fourier decomposition of the wave function gives
     an energy $E_{FT}= \tr(M)$. Generally, this spectral energy  differs from the energy obtained from the 
     variational calculation.
      However, the two energies must converge to the same value if we consider a sufficiently large number of 
     Slater determinants so that the exact wave function is sufficiently well approximated. This considerations 
     must  be kept in mind when we extract energies using spectral decomposition of the wave functions. 
     In general one can define the following spectral density of a Hamiltonian $\HH$ relative to some state $|\phi_0>$
$$
\rho(E)= <\phi_0|\del(E-\HH)|\phi_0>= \sum_n |<\phi_0|n>|^2 \del(E-E_n)
\eqno(41)
$$
     where $E_n$ are the energies for the eigenstates $|n>$. The spectral density can be obtained from the
     Fourier transform of the time correlation function $<\phi_0|\phi(t)>$, where $|\phi(t)>$ is obtained from
     the time evolution of the initial state $|\phi_0>$, as
$$
\rho(E)={1 \over \pi} Re \; lim_{\Gamma\rightarrow 0^+}  \int_0^{\infty} \exp^{i (E+i\Gamma)t} <\phi_0|\phi(t)>
\eqno(42)
$$
     The number of the Slater determinants has to be sufficiently large for this method to be reliable. 
     Moreover, if the initial state is the static solution of the imaginary time evolution, we would obtain
     only one pole corresponding to $E=\tr(M)$, which is obviously wrong in the HF case. Hence eq. (42)
     gives reasonable estimates for the eigenvalues only if there is reasonable fragmentation of 
     $\rho(E)$ for a sufficiently large number of Slater determinants. Moreover, for this method to be reliable one has
     to show that the spectrum obtained in this way, is independent of the initial wave function $|\phi_0>$.
     In this work we will not study the convergence properties of this method. We prefer to obtain the 
     static energies using variational methods or by imaginary time propagation since we obtain upper bounds
     for the energy, while the energies obtained with eq. (42) are not upper bounds to the exact values.
\par
     As discussed in the next sections in the contest of the boost method for strength functions, 
     we need static solutions to a high degree
     of accuracy. The reason is the following. The expectation values of a one-body operator $Q=\ad_i q_{ij} a_j$ 
     are given by
$$
<\psi(t)|Q|\psi(t)>= \sum_{S,S'} \det( V  U) \tr(\rho q)
\eqno(43)
$$
     In the static limit, since neither the determinant nor $\rho$ change in time, $<\psi(t)|Q|\psi(t)>$ is constant. This is 
strictly
     true if we have determined the exact variational wave function. A small error in these wave functions can give
     rise to a spurious time dependence in the expectation values. However, the purpose of the boost method is
     to perturb slightly the variational wave function  with a boost of the type $ \exp^{i\eta Q}$, for small values of
     $\eta$, and  to analyze the time dependence of the expectation values of $Q$ in order to obtain the strength
     function.  
     We found that very small changes in the 
     energies of the ground state as we proceed in the variational calculation, is not a good criterion. We prefer to use 
     the  fact that one-body
     observables should not change during the real time evolution if we have determined the static solution with sufficient 
     accuracy. This problem is hardly seen for a small number of Slater determinants since in these cases it is not difficult to 
     determine the static solutions with the necessary accuracy. 
      This criterion is essential, especially for small values of $\eta$.
\bigskip
\section{ A brief description of the numerical method.}
\bigskip
      We solve numerically EOM2 (eq.21) and  eq.(22b)) for $\dot V$.  
      As pointed out in the previous section, it is not advisable to store the 
      matrix $L_2$.
      However we can easily evaluate $L_2 v$ where $v$ is any vector. Actually, we can easily evaluate any 
      power of
      $L_2$ applied to $v$. The linear system of eq.(22b) can  be solved by projecting eq.(22b) into
      the subspace (known as Krylov subspace) generated by the vectors $ v, L_2v, (L_2)^2 v,..$, where $v$ is an arbitrary 
      trial solution of
      the linear system, followed by Gram-Schmidt orthonormalization.
      Since $L_2$ is Hermitian, its projection in the Krylov subspace  gives a tridiagonal matrix (just as in the 
      shell model method) and the linear system can then be efficiently solved. We have implemented the
      so called direct Lanczos method, the full detail of which (including the algorithm) can be found in ref. [13].
      With  this method the tridiagonal linear system is solved efficiently. 
      In our computer program the iterations stop when the residual vector $-iL_2\dot V- R_2$ has
      a norm less than $10^{-7}\div 10^{-11}$. The dimension of the Krylov subspace is less
      than the dimension of the linear system of eq.(22b) and, although it is advisable to implement 
      eq.(9), we found no actual need. 
\par
      In this work we considered ${}^6Li$ with the interaction given
      by the N3LO nucleon-nucleon potential renormalized using the Lee-Suzuki method to $4$  major
      harmonic oscillator shells.
      We considered $\hbar\om=12MeV$ and we added to the Hamiltonian the  center of mass
      Hamiltonian $\beta(P^2_{cm}/2mA+m A \om R^2_{cm}/2-3\hbar\om/2)$ with  $\beta=1$.
      The ground-state energies in MeV's as a function of the number of Slater determinants 
      are the following: $E(N_w=1)=-15.933, E(N_w=15)=-22.911, E(N_w=25)=-24.033$.
      Without using projectors to good quantum numbers,
      the absolute values of the energies converge slowly as a function of the number of 
      Slater determinants. For $N_w=200$, re-projecting the wave functions to good angular momentum and parity
      we obtained $E_{gs}=-28.672$. Since for large $N_w$ we have $E(N_w)\approx E_0 +const/N_w$ we
      can extract an extrapolated ground state energy of $-28.774$.

      Once $\dot V$ has been determined, we solve the differential equation in time
      using a rank-$4$ Runge-Kutta method. The time step used in the real time evolution is typically 
      $10^{-3}\div 10^{-4} MeV^{-1}$. In fig. 1 
      we show 
      the errors in the energies and in the overlaps $DE= E(t)-E(0)$ and 
      $DO=\ln( \CO(t)/\CO(0))$ as a function of the real time.    
\renewcommand{\baselinestretch}{1}
\begin{figure}
\centering
\includegraphics[width=10.0cm,height=10.0cm,angle=0]{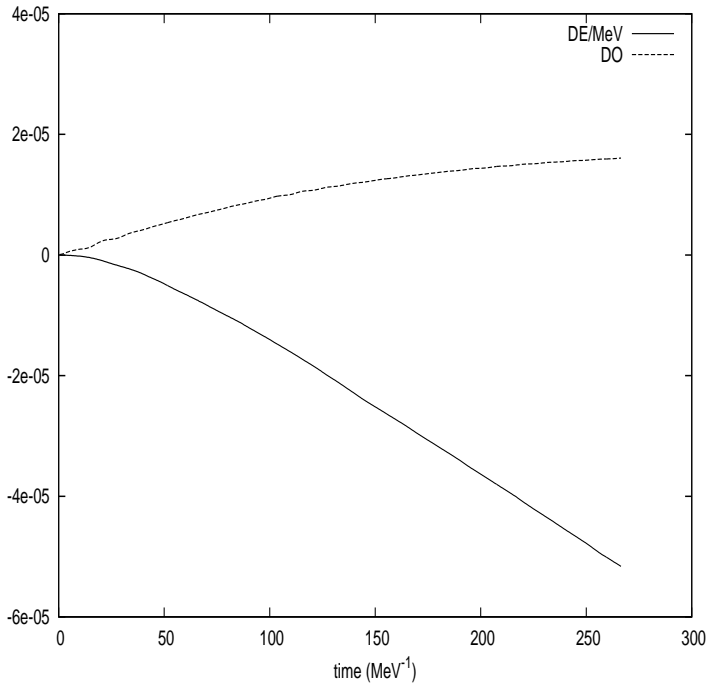}
\caption{Variation of the energy (in MeV) and variation of the norm as a function of time.
We took 3 Slater determinants with $4$ major shells at $\hbar\om=12MeV$ Snapshots are taken every $0.5 
MeV^{-1}$.}
\end{figure}
\renewcommand{\baselinestretch}{2}
\par\noindent
      In fig. 1 we took snapshots every $500$  time steps. Typically we ran the time evolution up to 
      $\simeq 50\div 100 MeV^{-1}$. The number of Lanczos iterations needed to converge depends on the number of Slater 
      determinants.
      For $1$ Slater determinant (TDHF) we need about $4$ Lanczos iterations to reach machine accuracy. This 
      number increases
      as we increases the number of Slater determinants. For example for $3$ Slater determinants we need 
      typically $36$
      Lanczos iterations and for $5$ Slater determinants we need about $50$ iterations.
     We perform only a sporadic check of the energy of the center
     of mass in order to ensure that the wave function does not develop spurious center of mass excitations.
     We verified that $E_{cm}(t)\approx 3\hbar\om/2$.
     Before leaving this section, let us make a few comments about the computer implementation of the method.
     In exact arithmetic, the Lanczos method will generate an orthogonal basis. With finite numerical
     accuracy, orthogonality is lost, preventing numerical convergence. Hence it is very important
     to re-orthogonalize the Krylov basis, as done in the shell model method.
     Let us briefly recall that if we have $n$ orthonormal vectors $|v_1>,..,|v_n>$ and we wish to add to this set
     another orthogonal vector $|v_{n+1}>$ starting from a vector $|u>$ we can use the so called classical
     Gram-Schmidt method, that is we evaluate $|u'>=|u>-\sum |v_k><v_k|u>$. In this case the  scalar
     products $<v_k|u>$ for $k=1,..,n$ can be evaluated independently using different processors. This classical Gram-Schmidt 
     method  however is known to be  numerically unstable for a large number of vectors.
      This instability can however be cured by first orthogonalizing
     $|u>$ to $|v_1>$, then the result is orthogonalized to $|v_2>$ and so on. This latter method is known as the modified
     Gram-Schmidt method and it is numerically stable. In this case, however, we cannot evaluate the several scalar products using
     different processors, since it is a sequential chain of calculations. 
     The instability of the classical Gram-Schmidt method can be bypassed  by simply repeating 
     two or three times the orthogonalization procedure. We have implemented both the iterated classical and the
     modified Gram-Schmidt re-orthogonalization in the direct Lanczos method. 
\bigskip
\bigskip
\section{ Strength functions.}
\bigskip
      We evaluate the strength function for a one body operator $\HQ$
$$
S(E)=\sum_n |<n|\HQ|0>|^2 \del(E-E_n^*)
\eqno(44)
$$
     $E_n^*$ being the excitation energy of the n-th  eigenstate, with the boost method, as
     follows. First we determine the   ground-state of the system $|0>$, then at 
     time $t=0^+$   we boost the system with the unitary operator
$$
|\psi(0^+)=\expo(i \eta \HQ)|0>
\eqno(45)
$$
      where $\HQ$ is a one-body operator. For sufficiently small values of the parameter $\eta$,  only linear terms in 
      $\eta$ can be 
      retained. We then evolve this wave function in time by solving the equations of 
      motion  EOM2 and evaluate
$$
Q(t) = <\psi(t)|\HQ|\psi(t)>-<\psi(0^+)|\HQ|\psi(0^+)>
\eqno(46)
$$
     The strength function can then be obtained using the Fourier transform of $Q(t)$ (see for example ref. [22])
$$
\overline Q(E) = \int_0^T dt \exp^{i (E+i\Gamma)t} Q(t)
\eqno(47)
$$
     for sufficiently large $T$ such that $\exp^{-\Gamma T}$ is negligible via the relation
$$
S(E)={1\over \eta \pi} Im(\overline Q(E) )
\eqno(48)
$$
     Alternative methods for the determination of  strength functions can be found in ref. [23]
     and in ref. [24]. 
\par
     In eqs. (45) and (46) the ground-state is replaced by the static solution evaluated with high accuracy. 
     Only in this case
     we can safely  guarantee that the response of eq.(46) is proportional to $\eta$.
\renewcommand{\baselinestretch}{1}
\begin{figure}
\centering
\includegraphics[width=10.0cm,height=10.0cm,angle=0]{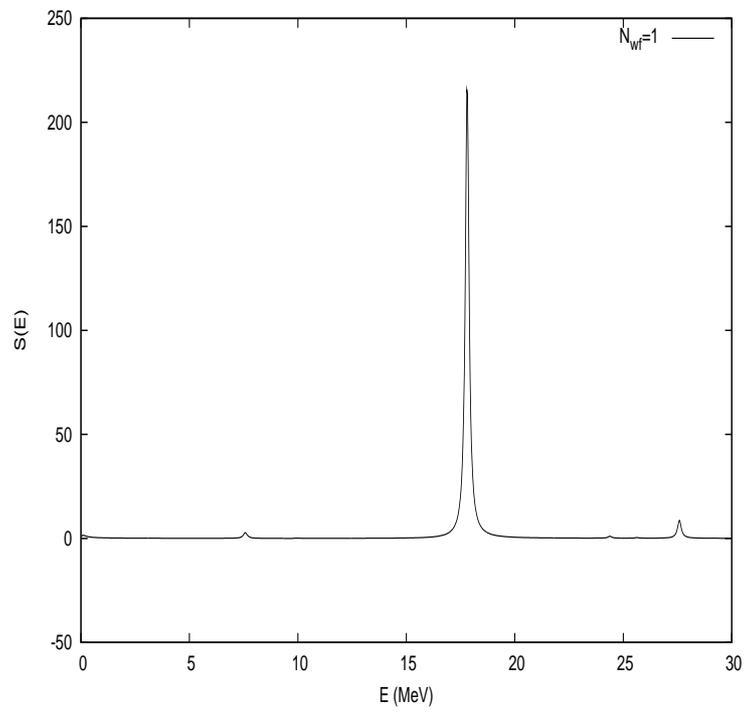}
\caption{Monopole strength function for $N_w=1$ (TDHF).}
\end{figure}
\renewcommand{\baselinestretch}{2}
\renewcommand{\baselinestretch}{1}
\begin{figure}
\centering
\includegraphics[width=10.0cm,height=10.0cm,angle=0]{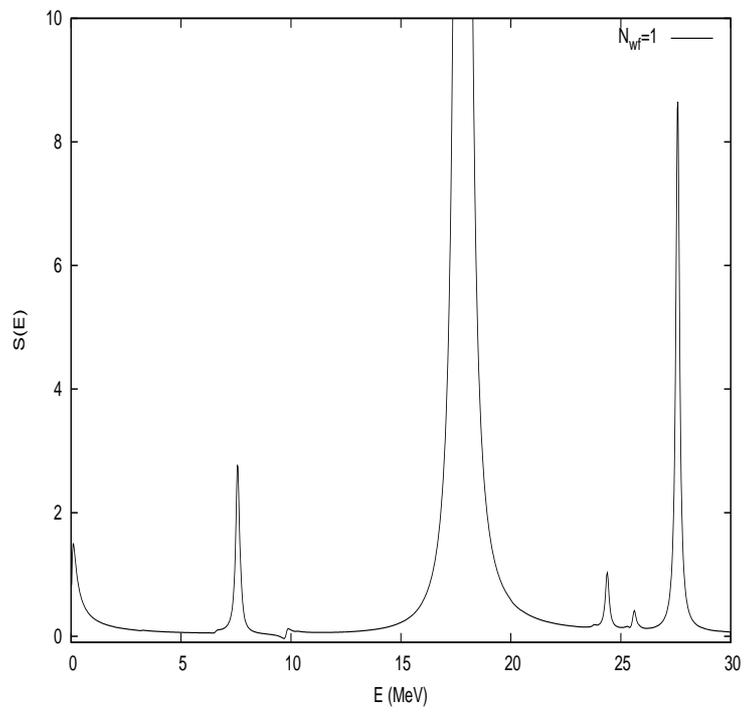}
\caption{Magnification of fig. 2.}
\end{figure}
\renewcommand{\baselinestretch}{2}
\par\noindent
      The width $\Gamma$ is small and such that  very high frequency oscillations in the Fourier transform 
      are smoothed out. Typically We take
       $\Gamma=0.1 MeV$ since we would like to resolve also discrete levels. If we are interested
      only in giant resonances we can afford much larger values.
       We need to evolve the system after the boost for about $T=50\div 100 MeV^{-1}$.
      In these exploratory calculations we have used the isoscalar monopole operator $Q=r^2$.
\renewcommand{\baselinestretch}{1}
\begin{figure}
\centering
\includegraphics[width=10.0cm,height=10.0cm,angle=0]{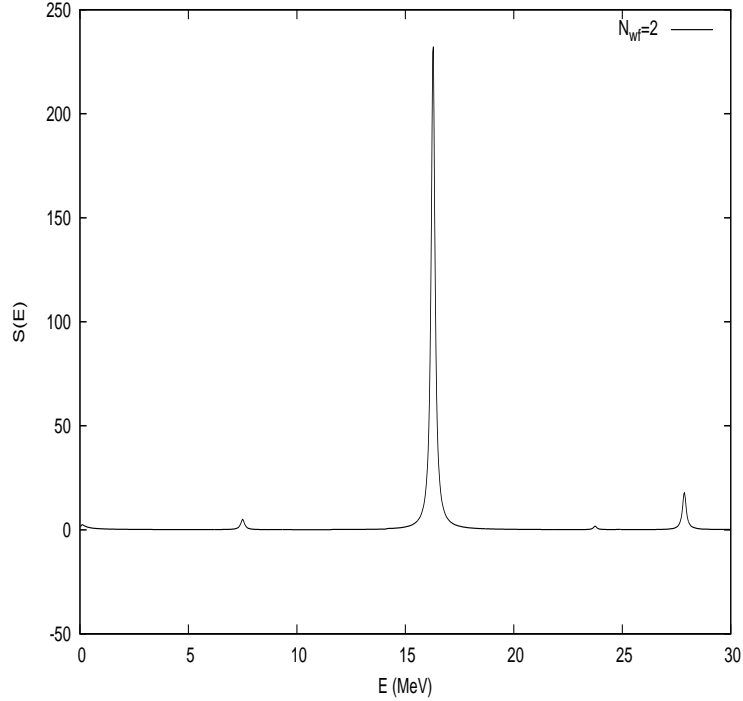}
\caption{Monopole strength function for $N_w=2$ Slater determinants.}
\end{figure}
\renewcommand{\baselinestretch}{2}
\renewcommand{\baselinestretch}{1}
\begin{figure}
\centering
\includegraphics[width=10.0cm,height=10.0cm,angle=0]{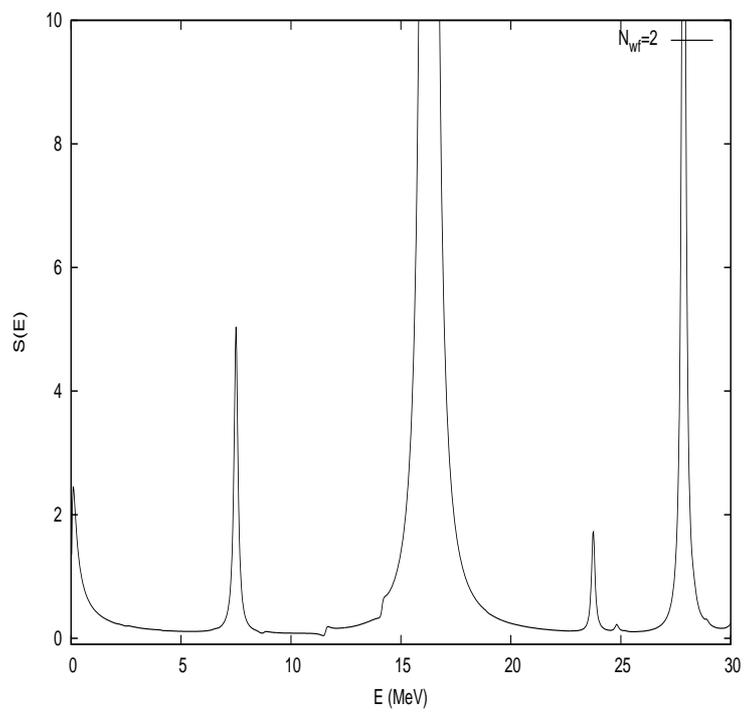}
\caption{Magnification of fig. 4.}
\end{figure}
\renewcommand{\baselinestretch}{2}
\par\noindent
      In figs. (2) and (3) we show the results obtained in the case of  one Slater determinant (TDHF). Fig (3) is a 
      magnification 
      of fig (2). The strength function is dominated by the dominant peak at $E\simeq 17.8MeV$. Some weaker peaks
      can be seen at $ E\simeq 0.1MeV$, $E\simeq 7.56 MeV$ and $E\simeq 27.6 MeV$. 
\renewcommand{\baselinestretch}{1}
\begin{figure}
\centering
\includegraphics[width=10.0cm,height=10.0cm,angle=0]{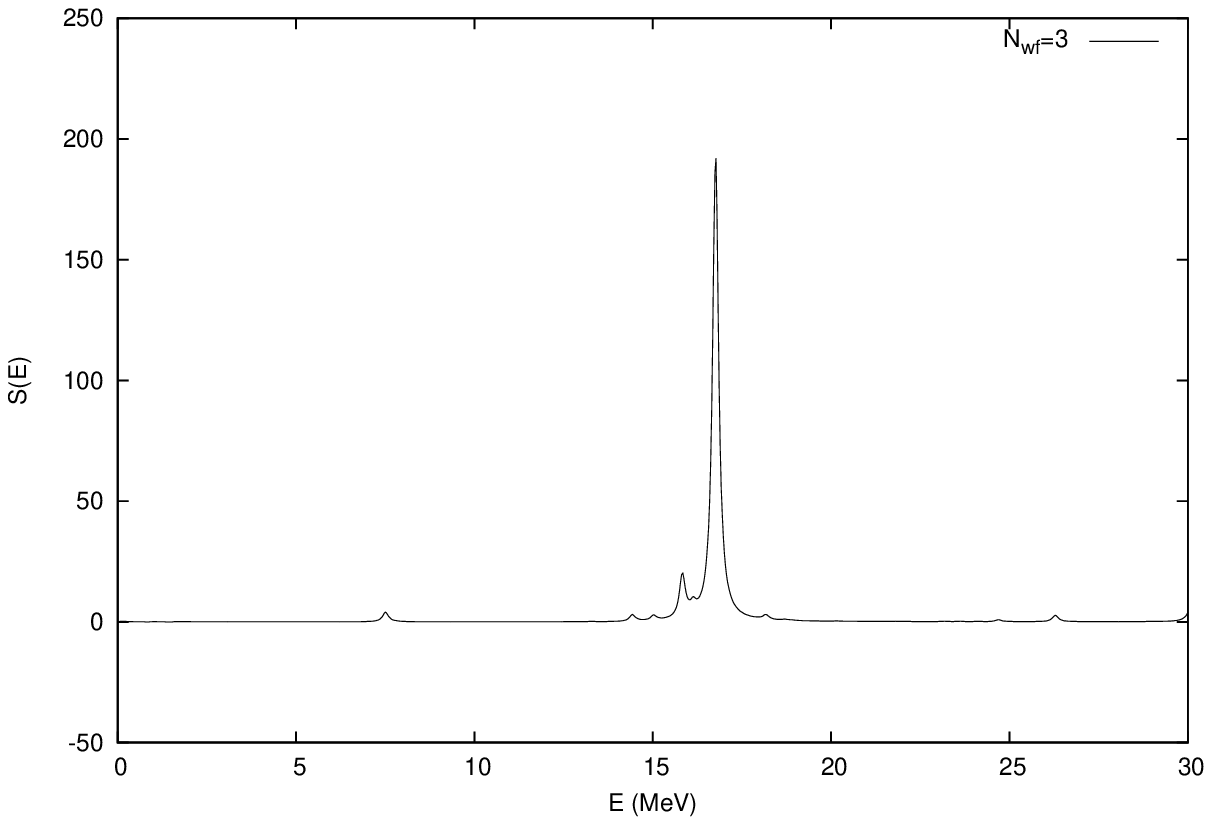}
\caption{Strength function for $N_w=3$ Slater determinants.}
\end{figure}
\renewcommand{\baselinestretch}{2}
\renewcommand{\baselinestretch}{1}
\begin{figure}
\centering
\includegraphics[width=10.0cm,height=10.0cm,angle=0]{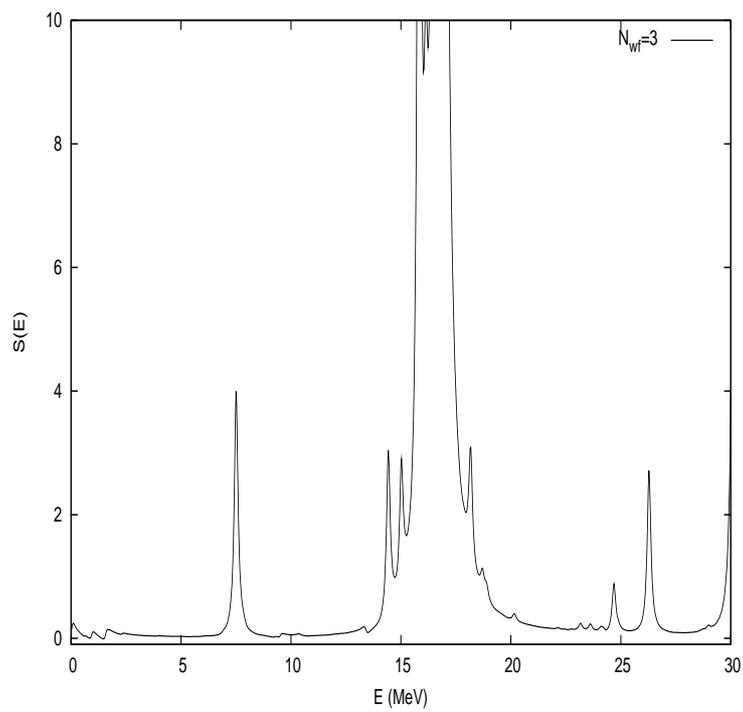}
\caption{Magnification of fig. 6.}
\end{figure}
\renewcommand{\baselinestretch}{2}
\par\noindent
      With $2$ Slater determinants we obtained the results shown in figs. (4) and (5).The dominant peak
      is now at $E\simeq 16.2 MeV$. The secondary maxima are at $E\simeq 0.1 MeV, 7.5 MeV, 27.8 MeV$, almost on the same 
      position of the TDHF case.
\renewcommand{\baselinestretch}{1}
\begin{figure}
\centering
\includegraphics[width=10.0cm,height=10.0cm,angle=0]{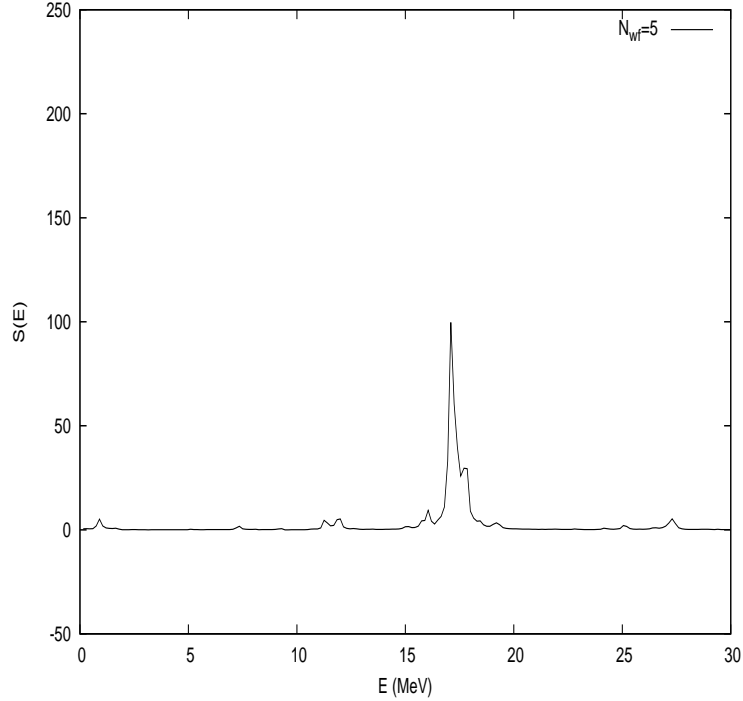}
\caption{Strength function for $N_w=5$ Slater determinants.}
\end{figure}
\renewcommand{\baselinestretch}{2}
\renewcommand{\baselinestretch}{1}
\begin{figure}
\centering
\includegraphics[width=10.0cm,height=10.0cm,angle=0]{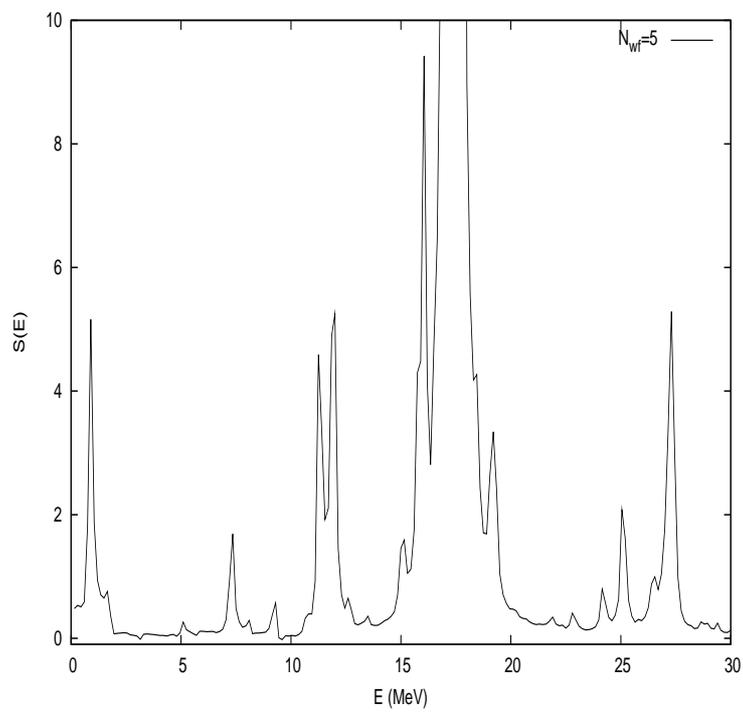}
\caption{Magnification of fig. 8.}
\end{figure}
\renewcommand{\baselinestretch}{2}
\par\noindent
      With $3$ Slater determinants, we obtained the results of figs, (6) and (7). The main peak at $E\simeq 16.7 MeV$
      shows considerable fragmentation around $15$ MeV while the secondary peak at $7.5 MeV$ is nearly unchanged.
      The peak around $27$ MeV has nearly disappeared and has moved to lower excitation energies.
      Similar plots, using $5$ Slater determinants, are shown in figs. (8) and (9),
      We also considered a larger number of Slater determinants, although for smaller values of $T$, $N_w=15$ and $N_w=25$.
      In these latter cases, some high frequency oscillations still remain. The results for the strength functions are shown 
      in figs. (10) and (11). Note that the structure of the strength function has changed considerably, pointing
      out to the need to consider a larger number of Slater determinants.
\renewcommand{\baselinestretch}{1}
\begin{figure}
\centering
\includegraphics[width=10.0cm,height=10.0cm,angle=0]{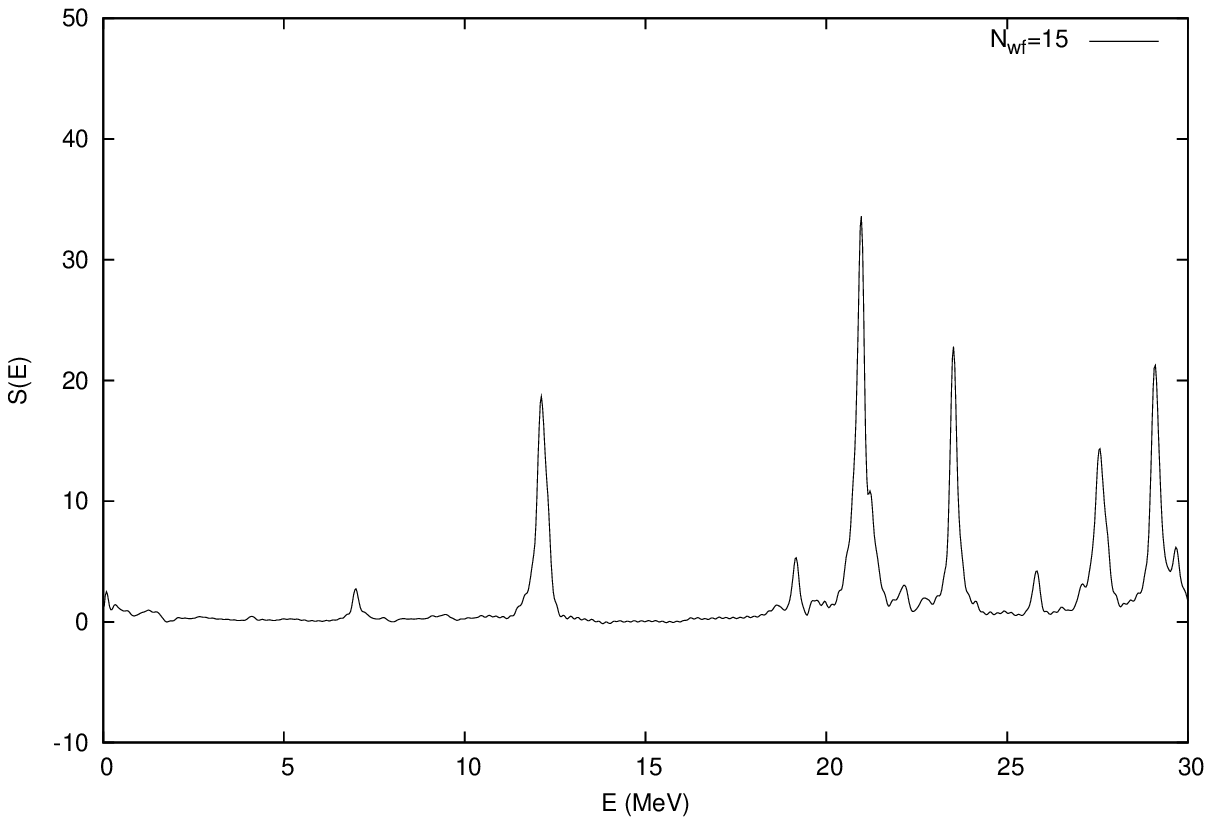}
\caption{Strength function for $N_w=15$ Slater determinants.}
\end{figure}
\renewcommand{\baselinestretch}{2}
\renewcommand{\baselinestretch}{1}
\begin{figure}
\centering
\includegraphics[width=10.0cm,height=10.0cm,angle=0]{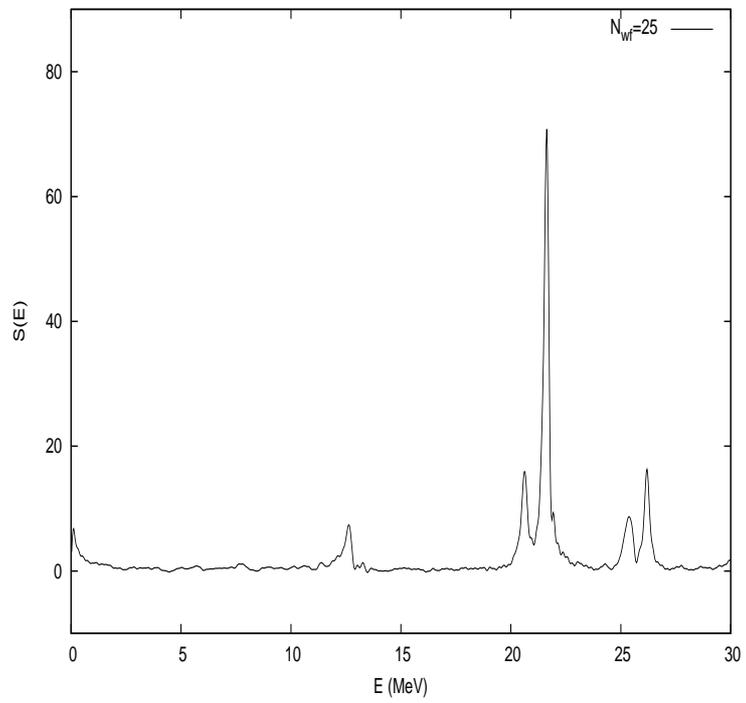}
\caption{Strength function for $N_w=25$ Slater determinants.}
\end{figure}
\renewcommand{\baselinestretch}{1}
\par
      We have not studied the strength function for a larger number of major shells and as a function of
      the harmonic oscillator frequency for increasing number of Slater determinants. Such a study is necessary in order
      to promote the TDMD method as an ab-initio method.
      The Lee-Suzuki renormalization method in harmonic oscillator space gives a Hamiltonian which depends on the
      number of particles, on the number of major harmonic oscillator shells and on the harmonic oscillator
      frequency. Therefore it is  a priori difficult to guess what would be the effect on the strength
      function of a larger number of harmonic oscillator shells and a larger number of Slater determinants.
      As we increase the number of major shells, the interaction becomes "harder" at short distances and
      we expect, on general grounds, to need an increasing number of Slater determinants. Moreover, giant
      resonances lie in the continuum and, for a proper description of their width we need a large
      single-particle space. Differently stated, if we select large single-particle basis the interaction
      becomes stronger. Part of these problems can be alleviated using low momentum interactions,
      whereby the NN interaction is renormalized in momentum space and does not depend on the number
      of oscillator shells (cf. for example ref. [25] and references in there). 
      Moreover, we expect on general grounds that large values of the harmonic oscillator frequency would
      give peaks further apart. Small values of $\hbar\Om$ should give a better approximation
      to the continuum giving a smaller distance among the peaks of the strength function. To some extent,
      a simple remedy to the lack of the continuum is to increase the width $\Gamma$.
\renewcommand{\baselinestretch}{1}
\begin{figure}
\centering
\includegraphics[width=10.0cm,height=10.0cm,angle=0]{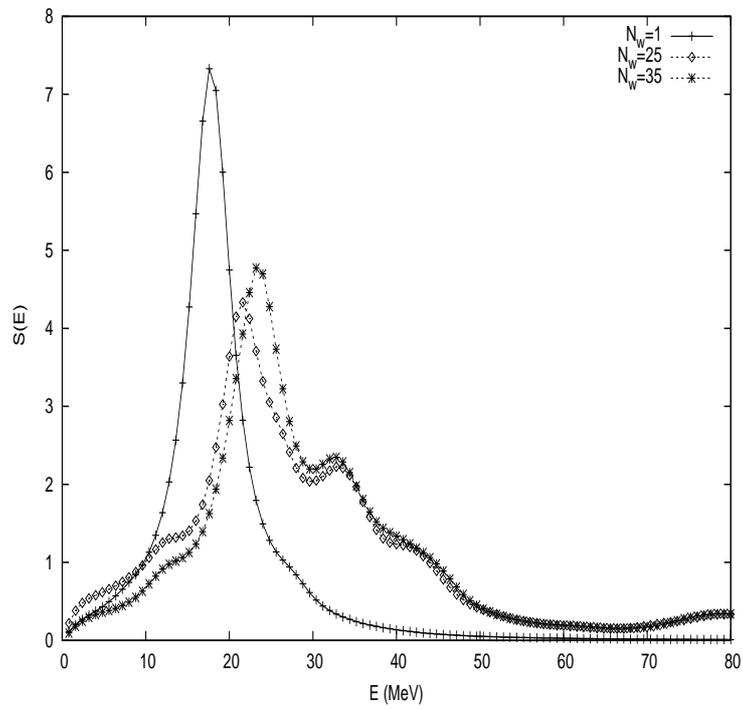}
\caption{Low resolution strength functions for $N_w=1,25,35$, for $\Gamma=3 MeV$.}
\end{figure}
\renewcommand{\baselinestretch}{1}
\par
      In fig. 12 we compare the monopole strength functions for $N_w=1,25,35$ evaluated with $\Gamma=3 MeV$.
      This comparison gives an idea, although with low energy resolution, of the degree of convergence
      as we increase the number of Slater determinants. Some discrepancy between $N_w=25$ and $N_w=35$
      still remains, but the shapes are very similar. The TDHF result, instead, is different. A possible cause
      of  the discrepancy between the TDHF strength and the ones for $N_w=25$ and $N_w=35$ is the angular
      momentum content of the wave functions. In this work we did not project the wave functions to
      good angular momentum. Since the Slater determinants break rotational symmetry we do not expect
      that the wave functions to have good angular momentum, especially for a small number of Slater
      determinants. We have checked the expectation values of $J^2$ for $N_w=1,25,35$. The results
      are the following: $<J^2>_{N_w=1}=6.94$, $<J^2>_{N_w=25}= 4.54$ and $<J^2>_{N_w=35}= 4.21$, instead
      of the exact value $<J^2>=2$. Let us recall that we are probing the system with a scalar probe.
      In the TDHF case, the initial wave function contains too many spurious components which are 
      excited by the monopole probe. For large numbers of Slater determinants these are smaller and
      the monopole strengths are very similar.
\par
      Although we do not have a formal proof, if we have a very large number of Slater determinants, 
      it is reasonable to assume that  the number of static solutions is equal to the dimension of the Hilbert space.
      The number of peaks in the strength function is equal to number of static solutions that can
      be connected by the excitation operator to the ground-state. Unfortunately we do not know
      the number of static solutions of the type of eq.(3) for a given $N_w$. In the derivation of the boost
      method it is tacitly assumed that the static solutions are eigenstates of the Hamiltonian.
\par
      For some recent works that take into account the continuum  in the TDHF and in the TDHFB approximations,
      see for example refs. [26],[27].
      Our main goal in this work is
      to define the time dependent method, solve the equations of motion and verify our computer programs. 
      More applications will be presented in future works.
\bigskip
\par
{\it{ Acknowledgments.}}
\par\noindent
      We acknowledge the CINECA award under the ISCRA initiative, for the availability of high performance computing 
      resources and support. We also acknowledge the use of computational resources at CILEA.

\vfill
\eject
\end{document}